\documentstyle[sprocl,epsfig,amsmath]{article}

\def\NPB{{\em Nucl. Phys.} B}

\def\PRL{\em Phys. Rev. Lett.}

\def\EPJ{{\em Eur. Phys. J.} C }

\def\be{\begin{equation}}
\def\ee{\end{equation}}
\def\bea{\begin{eqnarray}}
\def\eea{\end{eqnarray}}

\begin{document}

\thispagestyle{empty}
 
\begin{flushright}
{\large \tt LC-PHSM-2004-020}\\[20mm]
\end{flushright}
 
\begin{center}
    {\LARGE \bf Experimental Tools for SPA}\\[10mm]
    {\normalsize \sc Peter Wienemann}\\[5mm]
    {\normalsize \it Deutsches Elektronen-Synchrotron DESY,\\[0.2ex]
     Notkestr.~85, 22607 Hamburg, Germany}
\end{center}
 
\vspace{30mm}
 
\begin{abstract}
  \noindent
  Provided SUSY is realized in Nature, future colliders like the Large
  Hadron Collider (LHC) and a future $\text{e}^+\text{e}^-$ linear
  collider (LC) will provide a wealth of data on SUSY phenomena. One
  important task will be to extract the Lagrangian parameters at the
  electroweak scale from the numerous measured observables and to
  extrapolate them to a high scale to check whether unification takes
  place or to learn about the SUSY breaking mechanism. To accomplish
  such a task, two new programs, SFITTER and Fittino, have recently
  been developed. This talk introduces both programs and presents
  first results obtained with them.
\end{abstract}
 
\vspace{20mm}
 
\begin{center}
   {\normalsize Proceedings of the International Conference on Linear Colliders (LCWS 04)\\
                Paris, April 19-23, 2004 }
\end{center}
 
\newpage

\setcounter{page}{1}

\title{EXPERIMENTAL TOOLS FOR SPA\\}

\author{PETER WIENEMANN}

\address{Deutsches Elektronen-Synchrotron DESY,\\
         Notkestr.~85, 22607 Hamburg, Germany}


\maketitle\abstracts{
  Provided SUSY is realized in Nature, future colliders like the Large
  Hadron Collider (LHC) and a future $\text{e}^+\text{e}^-$ linear
  collider (LC) will provide a wealth of data on SUSY phenomena. One
  important task will be to extract the Lagrangian parameters at the
  electroweak scale from the numerous measured observables and to
  extrapolate them to a high scale to check whether unification takes
  place or to learn about the SUSY breaking mechanism. To accomplish
  such a task, two new programs, SFITTER and Fittino, have recently
  been developed. This talk introduces both programs and presents
  first results obtained with them.
}

\section{Introduction}
\label{sec:introduction}

Assuming Supersymmetry (SUSY) will be established experimentally,
future colliders like the Large Hadron Collider (LHC) and a future
$\text{e}^+ \text{e}^-$ linear collider will abundantly produce SUSY
particles. Huge data sets will allow precise measurements of SUSY
phenomena. Since theory parameters are in general not observables, one
of the important tasks will be to extract Lagrangian parameters from
the available measurements. This requires a mapping between
observables and theory parameters within a certain theory framework,
e.~g.~the Minimal Supersymmetric Standard Model (MSSM).

On tree-level some SUSY sectors depend only on a small number of
parameters (e.~g.~the chargino sector). Analytical relations can be
derived to easily calculate the low scale Lagrange parameters from the
measurements. On loop-level the situation is more complicated. In
principle every parameter depends on every observable and vice versa.
Therefore some elaborate method is needed to include loop corrections
in precise SUSY parameter determinations. The Supersymmetry Parameter
Analysis (SPA) Project \cite{ref:spa} has been set up to develop such
techniques.

\section{SFITTER and Fittino}
\label{sec:sfitterfittino}

\begin{figure}[ht] 
\begin{center}
    \epsfig{file=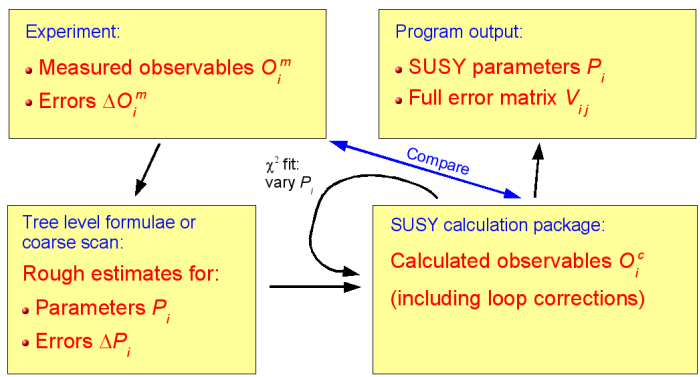,width=0.78\textwidth}
\end{center}
\caption{The iterative method used to extract the Lagrangian parameters
  from the measured observables.}
\label{fig:iterativeapproach}
\end{figure}

In the recently presented programs SFITTER \cite{ref:sfitter} and
Fittino \cite{ref:fittino}, an iterative approach has been chosen to
tackle the challenge described in Section \ref{sec:introduction}.
Starting from parameter start values obtained from a coarse scan (as
done in SFITTER) or tree-level formulae (as done in Fittino), the
corresponding observables are calculated (including loop corrections)
using a SUSY calculation package. The calculated observables are
compared to the measured ones.  Subsequently the Lagrangian parameters
are varied until the corresponding calculated observables agree with
the measurements. As fit output a set of SUSY parameters is obtained
including the full error matrix and, if requested, two-dimensional
uncertainty contours.  A sketch of the chosen iterative procedure is
shown in Fig.~\ref{fig:iterativeapproach}.

In their present versions SFITTER and Fittino make use of different
program components. SFITTER utilizes SUSPECT \cite{ref:suspect} for
mass predictions, MSMlib \cite{ref:msmlib} for branching ratios and
$\text{e}^+ \text{e}^-$ cross sections, and PROSPINO 2.0
\cite{ref:prospino} for pp cross sections. Fittino employs SPheno
2.2.0 \cite{ref:spheno} for masses, branching ratios and $\text{e}^+
\text{e}^-$ cross sections. Both programs use MINUIT \cite{ref:minuit}
to perform the fit. The communication between SFITTER/Fittino and the
various SUSY calculation packages takes place using the SUSY Les
Houches Accord (SLHA) \cite{ref:slha}. This ensures that the SUSY
calculation packages can be easily exchanged and thus allows to
cross-check the fit results without much effort.

\section{SPS1a Example Fits}

To verify the procedure described in Section~\ref{sec:sfitterfittino},
example fits have been carried out for the SPS1a scenario
\cite{ref:sps}.  Anticipated measurements of masses, cross sections
and branching ratios at LHC and a TeV linear collider serve as input
to various fits. A detailed list of used measurements can be found in
\cite{ref:lhclc} for the SFITTER fits (only masses are used as input)
and in \cite{ref:philipphd} for the Fittino results (both masses and
cross-sections serve as input).

\subsection{mSUGRA fit}

Assuming gravity mediated SUSY breaking (mSUGRA), the SUSY Lagrangian
is only determined by four high-scale parameters and the sign of the
Higgsino mass parameter $\mu$. These four parameters are the universal
scalar and gaugino masses $m_0$, $m_{1/2}$, the trilinear coupling
$A_0$ and the ratio of the two Higgs vacuum expectation values $\tan
\beta$. A fit of the four high-scale parameters has been performed
with SFITTER for three different input sets: ``LHC only'' and ``LC only''
measurements and a combined set using anticipated results from both
machines. $\text{sign}(\mu)$ has been fixed. The results for the three
different input sets are shown in Tab.~\ref{tab:msugrafit}.
\begin{table}[ht]
\scriptsize
\begin{center}
\begin{tabular}{l r |  r r | r r | r r}
    \hline
          & SPS1a & LHC & $\Delta_{\text{LHC}}$ & LC & $\Delta_{\text{LC}}$ & LHC+LC
          & $\Delta_{\text{LHC+LC}}$ \\
    \hline
    $M_0$ (GeV)     &  100 & 100.03 &  4.0 & 100.03 & 0.09 & 100.04 & 0.08 \\
    $M_{1/2}$ (GeV) &  250 & 249.95 &  1.8 & 250.02 & 0.13 & 250.01 & 0.11 \\
    $\tan \beta$    &   10 &   9.87 &  1.3 &   9.98 & 0.14 &   9.98 & 0.14 \\
    $A_0$ (GeV)     & -100 & -99.29 & 31.8 & -98.26 & 4.43 & -98.25 & 4.13 \\
    \hline
\end{tabular}
\end{center}
\normalsize
\caption{Results of the mSUGRA fit obtained with SFITTER $^{11}$.}
\label{tab:msugrafit}
\end{table}
As can been seen, the determination of the mSUGRA parameters are
dominated by the LC measurements. The contraints imposed by mSUGRA
unification bring about that the missing strongly interacting
particles at a LC have no influence on the precision of the parameter
determination.

\subsection{General MSSM fit}

Moreover fits have been performed in the general MSSM scenario which
does not make any assumptions on the SUSY breaking scenario. The price
for this generality is that one ends up with 105 SUSY parameters
(masses, phases, mixing angles). To make a fit feasible, it has been
assumed that there is no mixing between generations and within the
first two generations and that all phases vanish. Imposing these
contraints, 24 SUSY parameters remain. In addition unification in the
first two generation has been imposed. To take parametric
uncertainties of SUSY observables into account, the top quark mass has
been fitted simultaneously. The trilinear couplings $A_{\text{top}}$,
$A_{\text{bottom}}$ and $A_{\tau}$ are replaced by the mixing
parameters $X_{\text{top}} = A_{\text{top}} - \mu / \tan \beta$,
$X_{\text{bottom}} = A_{\text{bottom}} - \mu \tan \beta$ and
$X_{\tau} = A_{\tau} - \mu \tan \beta$ in order to reduce correlations
with $\tan \beta$. The outcome of a fit with Fittino under the above
assumptions is shown in Tab.~\ref{tab:generalmssmfit}.
\begin{table}[ht]
\scriptsize
\begin{center}
\begin{tabular}{l r r r | l r r r}
    \hline
                                    & SPS1a   & Value & Error &
                                    & SPS1a   & Value & Error \\
    \hline                
    $\tan \beta$                    &  10.0   &         10.0 &     0.3 &
    $\mu$ (GeV)                     & 358.6   &        358.6 &     1.1 \\
    $X_{\tau}$ (GeV)                & -3837   &        -3837 &    131  &
    $M_{\tilde{\text{e}}_R}$ (GeV)  & 135.76  &       135.76 &    0.39 \\
    $M_{\tilde{\tau}_R}$ (GeV)      & 133.33  &       133.33 &    0.75 &
    $M_{\tilde{\text{e}}_L}$ (GeV)  & 195.21  &       195.21 &    0.18 \\
    $M_{\tilde{\tau}_L}$ (GeV)      & 194.4   &        194.4 &    1.2  &
    $X_{\text{top}}$ (GeV)          & -506    &        -506  &    30  \\
    $X_{\text{bottom}}$ (GeV)       & -4441   &       -4441  &   1765   &
    $M_{\tilde{\text{d}}_R}$ (GeV)  & 528     &          528 &    18   \\
    $M_{\tilde{\text{b}}_R}$ (GeV)  & 524.7   &        524.7 &    7.7   &
    $M_{\tilde{\text{u}}_R}$ (GeV)  & 530     &          530 &    19   \\
    $M_{\tilde{\text{t}}_R}$ (GeV)  & 424.4   &        424.4 &    8.5   &
    $M_{\tilde{\text{u}}_L}$ (GeV)  & 548.7   &        548.7 &    5.2   \\
    $M_{\tilde{\text{t}}_L}$ (GeV)  & 500.0   &        500.0 &    8.1   &
    $M_1$ (GeV)                     & 101.81  &       101.81 &    0.06  \\
    $M_2$ (GeV)                     & 191.76  &       191.76 &    0.10  &
    $M_3$ (GeV)                     & 588.8   &       588.8  &    7.9   \\
    $m_{\text{A}}$ (GeV)            & 399.8   &       399.8  &    0.7   &
    $m_{\text{top}}$ (GeV)          & 174.3   &       174.3  &    0.3   \\
    \hline
\end{tabular}
\end{center}
\normalsize
\caption{Results of the general MSSM fit obtained with Fittino $^{12}$.}
\label{tab:generalmssmfit}
\end{table}
All fitted parameters agree well with the predicted values from SPheno
2.2.0. It was verified that the parameter uncertainties obtained from
the fit are properly estimated by looking at pull distributions from
fits with smeared observables for all fitted parameters.

\section{Conclusion}

An iterative procedure has been developed to determine SUSY Lagrangian
parameters from measured observables. The method has proven successful
in example fits for the SPA1a scenario. It was verified that both the
fit values and uncertainties have been properly estimated in these
fits. Therewith two powerful tools are available for SPA.

\end{document}